\documentclass[twocolumn,superscriptaddress,showpacs,aps,amsmath,amssymb]{revtex4}
\usepackage{bm}
\usepackage{amsmath,amssymb,bm}
\usepackage{graphicx}

\newcommand{\B}{\text{\scriptsize res}}

\newcommand{\T}{{\rm total}}

\newcommand{\nl}{\nonumber \\}

\newcommand{\ep}{\epsilon}
\newcommand{\w}{\omega}

\newcommand{\be}{\begin{equation}}
\newcommand{\ee}{\end{equation}}
\newcommand{\bea}{\begin{eqnarray}}
\newcommand{\eea}{\end{eqnarray}}
\newcommand{\bsube}{\begin{subequations}}
\newcommand{\esube}{\end{subequations}}
\newcommand{\Eq}[1]{Eq.\,\eqref{#1}}

\newcommand{\Fig}[1]{Fig.\,\ref{#1}}
\newcommand{\Figs}[1]{Figs.\,\ref{#1}}

\newcommand{\comments}[1]{}

\begin{document}

\title{Long-range Exchange Interaction in Triple Quantum Dots in the Kondo Regime}

\author{YongXi~Cheng}
\affiliation{Department of Physics, Renmin University of China, Beijing
100872, China }
\affiliation{Department of Science, Taiyuan Institute of Technology, Taiyuan
030008, China }

\author{YuanDong~Wang}
\affiliation{Department of Physics, Renmin University of China, Beijing
100872, China }

\author{JianHua~Wei}\email{wjh@ruc.edu.cn}
\affiliation{Department of Physics, Renmin University of China,
Beijing 100872, China }

\author{ZhenGang Zhu}
\affiliation{School of Electronic, Electrical and Communication Engineering,
 University of Chinese Academy of Sciences, Beijing 100049, China }

\author{YiJing~Yan}
\affiliation{Hefei National Laboratory for Physical Sciences at the Microscale,
University of Science and Technology of China, Hefei, Anhui 230026, China}
\affiliation{ Department of Chemistry, Hong Kong University of Science and Technology, Kowloon, Hong Kong}

\date{\today}

\begin{abstract}
Long-range interactions in triple quantum dots (TQDs) in Kondo regime are investigated by accurately solving the three-impurity Anderson  model. For the occupation configuration of $(N_{1},N_{2},N_{3})=(1,0,1)$,  a long-range antiferromagnetic exchange interaction ($J_{\rm AF}$) is demonstrated and induces a continuous phase transition  from the separated Kondo singlet (KS) to the long-range spin singlet (LSS) state between edge dots. The expression of $J_{\rm AF}$ is analytically derived and numerically verified, according to which  $J_{\rm AF}$  can be conveniently manipulated via gate control of the detuning energy. The long-range entanglement of Kondo clouds are proved to be quite robust at strong inter-dot coupling limit. Under equilibrium condition, it induces an unexpected peak in the spectral function of the middle dot whose singly occupied level keeps much higher than the Fermi level. Under nonequilibrium condition, higher inter-dot tunneling barrier induces an anomalous enhancement  of current. These novel features can be observed in routine experiments.
\end{abstract}


\maketitle

\section{Introduction}
\label{Intro}

Long-range interaction as a high-order interaction originates from the superpositions of indirectly coupled states. It plays an important role in many-body physics and quantum computing \cite{2014nature198,2015prb035403,2013nn432, 2014prb161402}. For the latter, the long-range interaction makes it possible to manipulate distant quantum gate or qubit in one step, which is of higher operating efficiency and fault-tolerant capability than nearest-neighbor control in exchange-based quantum gates \cite{1998pra120}.
The triple quantum dots (TQDs) device provides an ideal platform for investigating the quantum manipulation \cite{2006prl036807,2007prb075306,2010prb075304,2008prb193306,2008apl013126,2009apl092103}. The long-range transport in serially coupled TQDs has been observed in recent experiments \cite{2013nn261,2013nn432,2014prl176803}. For example, Platero {\it et al.} measured a resonant transport line (in the area of the bipolar spin blockade) between the edge dots, which  suggests a long-range coherent superposition near the degenerate point of  $(N_{1},N_{2},N_{3})=(1,1,1)/(2,1,2)$ ($N_i$ is the number of electrons in $i$-th QD) \cite{2013nn261}. Shortly afterwards, the same group reported a long-range spin transfer near another degenerate point of  $(1,0,1)/(2,0,2)$, where QD2 keeps unoccupied during the tunneling process \cite{2014prl176803}.  Another group of Vandersypen {\it et al.}  demonstrated a high-order coherent tunneling between QD1 and 3 near the degenerate point of  $(0,1,0)/(1,1,1)$ through the observation of Landau-Zener-St\"{u}ckelberg interference \cite{2013nn432}.

In order to produce measurable current, all of above experimental results are achieved in the boundary of Coulomb blockade near degenerate points in the stability diagram. However, these regimes are not suitable for theoretical analysis of the long-range interaction (especially the long-range spin correlation or exchange interaction), since occupation numbers and magnetic moments of QD1 and QD3 are not conserved during the transport under bias in those boundaries. One better choice is to push the range of study deeply into the Coulomb blockade region far away from the degenerate points, such as the local moment regime of QD1 and QD3 where both the occupation number and spin are well defined. In order to produce measurable current or other observable features, we investigate the long-range exchange interaction and its effects in the Kondo regime.

The Kondo phenomenon itself is an important and interesting issue in TQDs. It results from the screening of a  localized spin by the delocalized spins from reservoirs (or leads), which presents a pronounced zero-bias conductance peak at temperatures below the Kondo temperature in QD systems, with a Kondo singlet (KS) formed \cite{1964ptp37,Ng881768}. Recently considerable theoretical efforts have been made in the topic of serial TQDs, such as the equilibrium and nonequilibrium Kondo transport properties \cite{2005prb045332}, Fermi-Liquid versus non-Fermi-liquid behavior \cite{2007prl047203}, and two-channel Kondo physics \cite{2005el218}. In addition, the Kondo phenomenon in other structures of TQDs have been discussed as well, including
the mirror symmetry TQDs \cite{2013prl047201,2006prb235310,2010prb165304}, triangular TQDs \cite{2009prb155330,2010prb115330,2011prb205304,2013prb035135}, and parallel TQDs \cite{2007prb115114}. To the best of our knowledge, none of these works concern the long-range exchange interaction and its effect on the Kondo phenomenon in TQDs.

\begin{figure}
\includegraphics[width=0.7\columnwidth]{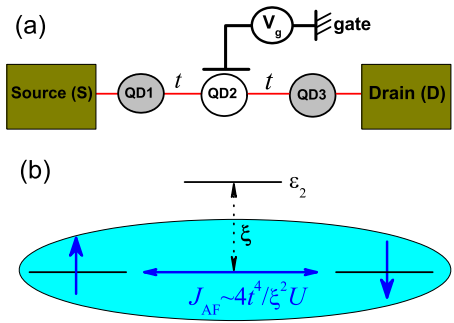}
\caption{ (a) The schematic diagram of the triple quantum dots (TQDs) system. In present work, QD1 and QD3 are symmetric and both in the localized momentum regime with $N_1=N_3=1$. The QD2  is nearly unoccupied with a gate-modulated on-site energy, $\varepsilon_2=-U/2+eV_g$. (b) The schematic diagram is shown for the long-range  antiferromagnetic exchange interaction ($J_{\rm AF}$) between QD1 and QD3 via high-order tunneling processes.}
\label{fig1}
\end{figure}

In present work, we study the long-range exchange interaction between QD1 and 3 in the Kondo regime
in  serial TQDs, by accurately solving the three-impurity Anderson  model with the hierarchical equation of motion (HEOM) formalism
\cite{Jin08234703,2012prl266403}.  The geometry is depicted in \Fig{fig1}(a). Two symmetrical edge dots (QD1 and 3) are in the local magnetic moment regime ($N_1=N_3=1$), and are coupled to the source (S) and drain (D) reservoir but decoupled from each other ($t_{13}=0$). The intermediate one (QD2) symmetrically couples to the QD1 and QD3 ($t_{12}=t_{23}=t$) via a variable singly-occupied level $\varepsilon_2$ modulated by a gate voltage $V_g$. In order to highlight the long-range correlation, we focus on the  occupation configuration of  $(N_{1},N_{2},N_{3})=(1,0,1)$ by pushing $\varepsilon_2$ high enough, as schematically shown in \Fig{fig1}(b).  In the limit of $(1,1,1)$, we have reported a reappearance of the Kondo phenomenon and worked out an effective ferromagnetic exchange interaction between QD1 and 3 \cite{2015epl17550}. In present work, as schematically shown in \Fig{fig1}(b), we will demonstrate a long-range antiferromagnetic exchange interaction ($J_{\rm AF}>0$), which can be simply expressed in terms of
\begin{equation}\label{JAF}
J_{\rm AF}\sim\frac{4t^{4}}{\xi^{2}U}
\end{equation}
where $\xi\equiv\varepsilon_2-\varepsilon_{1}$ ($\varepsilon_1$ being the on-site energy of QD1) is called the detuning energy, and $U$ ($U_i=U;i=1,2,3$)  is the on-dot Coulomb interaction. The effect of $J_{\rm AF}$ on Kondo features including spectral characters and Kondo current in TQDs will be discussed as well.

\section{Model and theory}
\label{Mod}

The total Hamiltonian for the system is described by the three-impurity Anderson model
\begin{align}\label{ha}
   H=H_{\rm dots}+H_{\rm res}+H_{\rm coup},
\end{align}
where the isolated TQD part is
\begin{align}\label{HS3}
   H_{\rm dots}=\sum_{\sigma, i=1,2,3}[\epsilon_{i\sigma}\hat{a}^\dag_{i\sigma}\hat{a}_{i\sigma} + U n_{i\sigma}n_{i\bar{\sigma}}]
 \nl    + t\sum_{\sigma}(\hat{a}^\dag_{1\sigma}\hat{a}_{2\sigma}+\hat{a}^\dag_{2\sigma}\hat{a}_{3\sigma}+\text{H.c.}),
  \end{align}
with $\hat{a}_{i\sigma}^\dag$ ($\hat{a}_{i\sigma}$) being the operator that creates (annihilates) a spin-$\sigma$ electron with energy $\epsilon_{i\sigma}$ in $i$-th QD. $n_{i\sigma}=\hat{a}^\dag_{i\sigma}\hat{a}_{i\sigma}$ is the operator of occupation number.

In what follows, the symbol $\mu$ is adopted to denote the electron orbital (including spin, space, \emph{etc.}) in the system for brevity, i.e.,  $\mu=\{{\sigma},i...\}$. The device reservoirs are treated as single-particle systems with the Hamiltonian as $H_{\rm res}=\sum_{k\mu\alpha={\rm S,D}}\epsilon_{k\alpha}\hat{d}^\dag_{k\mu\alpha}\hat{d}_{k\mu\alpha}$, with $\epsilon_{k\alpha}$ being the energy of an electron with wave vector $k$ in the $\alpha$-reservoir, and $\hat{d}^\dag_{k\mu\alpha}$($\hat{d}_{k\mu\alpha}$) corresponding creation (annihilation) operator for an electron with the $\alpha$-reservoir state $|k\rangle$ of energy $\epsilon_{k\alpha}$. The Hamiltonian of dot-reservoir coupling is $H_{\rm coup}=\sum_{k\mu\alpha}t_{k\mu\alpha}\hat{a}^\dag_{\mu}\hat{d}_{k\mu\alpha}+\text{H.c.}$ To describe the stochastic nature of the transfer coupling, it can be written in the reservoir $H_{\rm res}$-interaction picture as $H_{\rm coup}=\sum_{\mu}[f^\dag_{\mu}(t)\hat{a}_{\mu} +\hat{a}^\dag_{\mu}f_{\mu}(t)]$,
with $f^\dag_{\mu}=e^{iH_{\rm res}t}[\sum_{k\alpha}t^{*}_{k\mu\alpha}\hat{d}^\dag_{k\mu\alpha}]e^{-iH_{\rm res}t}$ being the stochastic interactional operator and satisfying the Gauss statistics. Here, $t_{k\mu\alpha}$ denotes the transfer coupling matrix element. The influence of electron reservoirs on the dots is taken into account through the hybridization functions, which is assumed Lorentzian form,
$\Delta_{\alpha}(\w)\equiv\pi\sum_{k} t_{\alpha k\mu}t^\ast_{\alpha k\mu} \delta(\w-\ep_{k\alpha})=\Delta W^{2}/[2(\w-\mu_{\alpha})^{2}+W^{2}]$, where $\Delta$ is the effective quantum dot-reservoir coupling strength, $W$ is the band width, and $\mu_{\alpha}$ is the chemical potentials of the $\alpha$-reservoir.

In this paper, the three-impurity Anderson model is accurately solved by the HEOM approach, which is established based on the Feynman-Vernon path-integral formalism with a general Hamiltonian, in which the system-environment correlations are fully taken into consideration \cite{Jin08234703,2012prl266403}.
The HEOM formalism is in principle accurate and applicable to arbitrary electronic systems, including Coulomb interactions, under the influence of arbitrary applied bias voltage and external fields. The outstanding issue of characterizing both equilibrium and nonequilibrium properties of a general open quantum system are referred to Refs. \cite{Jin08234703,2012prl266403,2015njp033009,2015epl17550,2015jcp234108,2014jcp054105}. It has been demonstrated that the HEOM approach achieves the same level of accuracy as the latest high-level numerical renormalization group and quantum Monte Carlo approaches for the prediction of various dynamical properties at equilibrium and nonequilibrium \cite{2012prl266403}.

The reduced density matrix of the quantum dots system $\rho^{(0)}(t) \equiv {\rm tr}_{\B}\,[\rho_{\T}(t)]$ and a set of auxiliary density matrices  $\left\{\rho^{(n)}_{j_1\cdots j_n}(t); n=1,\cdots,L\right\}$ are the basic variables in HEOM. $L$ denotes the truncated tier level. The equations governing the dynamics of open systems are in the form of \cite{Jin08234703,2012prl266403}:
\begin{align}\label{HEOM}
   \dot\rho^{(n)}_{j_1\cdots j_n} =& -\Big(i{\cal L} + \sum_{r=1}^n \gamma_{j_r}\Big)\rho^{(n)}_{j_1\cdots j_n}
     -i \sum_{j}\!     
     {\cal A}_{\bar j}\, \rho^{(n+1)}_{j_1\cdots j_nj}
\nl &
    -i \sum_{r=1}^{n}(-)^{n-r}\, {\cal C}_{j_r}\,
     \rho^{(n-1)}_{j_1\cdots j_{r-1}j_{r+1}\cdots j_n},
\end{align}
where ${\cal A}_{\bar j}$ and ${\cal C}_{j_r}$ are Grassmannian superoperators which are illustrated in detail in Refs.~\onlinecite{Jin08234703} and \onlinecite{2012prl266403}.

The dynamical quantities can be acquired via the HEOM-space linear response theory \cite{Wei2011arxiv}. The spectral function $A(\omega)$ exhibiting prominent Kondo signatures at low temperatures can be evaluated by a half Fourier transformation of correlation functions as
\begin{align}\label{hd}
    A_{\mu}(\omega)=\frac{1}{\pi}{\rm Re}\Big\{\int^{\infty}_{0}dt \{\tilde{\cal C}_{\hat{a}^\dag_{\mu}\hat{a}_{\mu}}(t)+
    [\tilde{\cal C}_{\hat{a}_{\mu}\hat{a}^\dag_{\mu}}(t)]^{\ast}\}e^{i\omega t}\Big\}.
  \end{align}
The electric current from $\alpha$-reservoir to system is given by
\begin{align}\label{hd}
    I_{\alpha}(t)=i\sum_{\mu}\mathrm{tr}_{s}[{\rho^\dag_{\alpha \mu}(t)\hat a_{\mu} -\hat a^\dag_{\mu}\rho^-_{\alpha \mu}(t)}],
  \end{align}
where $\rho^\dag_{\alpha \mu}=(\rho^-_{\alpha \mu})^\dag$ is the first-tier auxiliary density operator. The details of the HEOM formalism and the derivation of physical quantities are supplied in Refs.~\onlinecite{Jin08234703} and \onlinecite{2012prl266403}.

\section{Results and discussion}
\label{Rest}

\begin{figure}
\includegraphics[width=\columnwidth]{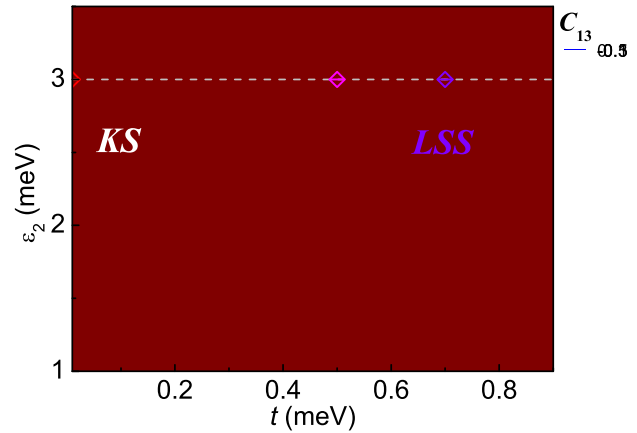}
\caption{ The spin-spin correlation function $C_{13}\equiv\langle \vec{S}_{1}\cdot \vec{S}_{3} \rangle-\langle \vec{S}_{1}\rangle\cdot\langle \vec{S}_{3}\rangle$ varies as a function of the on-site energy of QD2, i.e. $\varepsilon_2$, and the nearest inter-dot coupling strength $t$. Two phases are shown: the Kondo singlet (KS, $C_{13}\sim 0$) and long-range spin singlet (LSS, $C_{13}<0$). The  horizontal dashed line marks the gradual change of phases from KS to LSS, which is further elucidated by the spectral functions of the scatter points (see \Fig{fig3}). The  vertical dashed line marks the gradual change of phases with $\varepsilon_2$ at $t=0.7$ meV. The parameters are as follows: $U= 1.2$ (in unit of meV), $\epsilon_1= \epsilon_3= -0.6$, the bandwith of  reservoirs $W = 5.0$, the temperature $K_{B}T = 0.03$, and the hybridization width between  reservoirs and QDs is $\Delta = 0.3$.}
\label{fig2}
\end{figure}

As shown in \Fig{fig1}, we assume that QD1 and QD3 always keep electron-hole symmetry and their parameters are the same, in which $\varepsilon_{1}=\epsilon_{1}=-U/2$ and $\varepsilon_{3}=\epsilon_{3}=-U/2$. In order to figure out whether there may exist a long-range exchange interaction,  we calculate the spin-spin correlation function between QD1 and 3,
\begin{equation}\label{SSCF}
 C_{13}\equiv\langle \vec{S}_{1}\cdot \vec{S}_{3} \rangle-\langle \vec{S}_{1}\rangle\cdot\langle \vec{S}_{3}\rangle.
\end{equation}
In \Fig{fig2}, we depict $C_{13}$  as a function of the modulated on-site energy of QD2  ($\varepsilon_2=\epsilon_2+eV_g$) and the nearest inter-dot coupling strength $t$. The other parameters are as follows: the on-dot Coulomb correlation $U= 1.2$ (in unit of meV),  the bandwith of  reservoirs $W = 5.0$, the temperature $K_{B}T = 0.03$, and the hybridization widths between  reservoirs and QDs  $\Delta = 0.3$.
In present work, we set $\varepsilon_2>1.0$ meV to keep the configuration of $(1,0,1)$. Two phases are shown in the figure. The first one is the Kondo singlet (KS) which is characterized by near zero correlation between $\vec{S}_{1}$ and $\vec{S}_{3}$ ($C_{13}\sim 0$) at small $t$. The second one is the main finding of the present work called long-range spin singlet (LSS) characterized by finite $C_{13}$ ($C_{13}< 0$ in the figure), which proves that a long-range exchange interaction between $\vec{S}_{1}$ and $\vec{S}_{3}$ dose exist although direct coupling between them is absent. From the sign of $C_{13}$, we conclude that the long-range exchange is antiferromagnetic thus we suggest an effective interaction term as,
\begin{equation}\label{AFI}
 H_{13}=J_{\rm AF}\vec{S}_{1}\cdot \vec{S}_{3}.
\end{equation}
It is expected that $H_{13}$ plays an important role in quantum computing, which can expand the original idea of the exchange-based quantum gates \cite{1998pra120}, by manipulating distant quantum gate or qubit in one step.  We comment that the small value of $C_{13}$ in the bottom right corner of \Fig{fig2} results from the competition between the LSS phase and the effective ferromagnetic phase we have reported in Ref.~\onlinecite{2015epl17550}.

\begin{figure}
\includegraphics[width=\columnwidth]{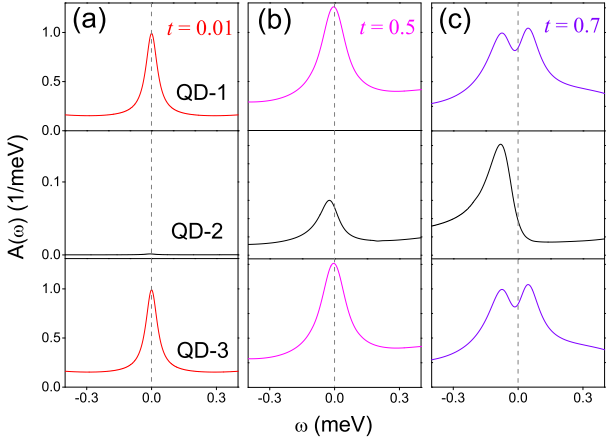}
\caption{(Color online) The spectral functions $A_i(\omega)$ of the TQDs are shown at $\varepsilon_2 = 3.0$ (in unit of meV) for different inter-dot coupling strengths: (a)  $t=0.01$, (b) $t=0.5$ and  (c) $t=0.7$. The top panel: $i=1$; the middle panel: $i=2$; and the bottom panel: $i=3$. The other parameters are the same as those in \Fig{fig2}.}
\label{fig3}
\end{figure}

More detailed information of the phase diagram in \Fig{fig2} can be illustrated by the spectral functions $A_{i\sigma}(\omega)$ in different phases. The spin-degeneracy makes $A_{i\uparrow}(\omega)=A_{i\downarrow}(\omega)=A_i(\omega)$ and the symmetry in our model also suggests $A_1(\omega)=A_3(\omega)$.  We select three characteristic points at $t=0.01$, 0.5 and 0.7 meV along the dashed line in \Fig{fig2} at $\varepsilon_2=3.0$ meV, and depict their corresponding $A_i(\omega)$ in \Fig{fig3}(a) to (c). By referring to \Figs{fig2} and \ref{fig3}, we find in the limit of weak inter-dot coupling ($t < 0.2$ meV),  the absence of long-range correlation ($C_{13}\sim0$) results in the individual screening of local momentums by the nearest  reservoirs, thus the degenerate KS state is formed.  The spectral function $A_1(\omega)/A_3(\omega)$ shows similar behavior to that in single QD with one Kondo peak at $\omega=0$, while $A_2(\omega)\sim0$ around $\omega=0$ due to the empty occupation, as shown in \Fig{fig3}(a). With the increase of the inter-dot coupling, the single peak of $A_1(\omega)/A_3(\omega)$ grows slightly higher due to the ``$t$-enhanced Kondo phenomenon" (figure not shown) \cite{2015njp033009}.

Further increasing $t$ to $t>0.3$ meV distinctly changes the Kondo features. As shown in \Fig{fig3}(b), the central peak of $A_1(\omega)/A_3(\omega)$  becomes much higher and wider at $t=0.5$ meV than that at $t=0.01$ meV, meanwhile  a small peak develops in $A_2(\omega)$ near $\omega=0$.  The latter is unusual, since the  QD2 is still empty and the emerging peak impossibly results from Kondo screening directly. The most possible mechanism is the long-range tunneling between QD1 and QD3 by the aid of $J_{\rm AF}$, or equivalently speaking, the electron wavefunctions  separately localized in QD1 and QD3 at $t\sim0$ becomes overlapping within QD2 now. At first glance, it is analogous to the ordinary double-well model in the textbook, however, what are localized in QD1 and QD3 here are not ordinary electrons but Kondo quasiparticles. It means that the Kondo quasiparticle in QD1 can tunnel to QD3 through QD2, via overlapping their wavefunctions which are nothing but the widely studied ``Kondo cloud'' \cite{2009arx0911}. Although we can not present the spatial distribution of Kondo clouds here,  we have demonstrated their long-range overlapping, or long-range quantum entanglement \cite{2015prl057203}.

\begin{figure}
\includegraphics[width=\columnwidth]{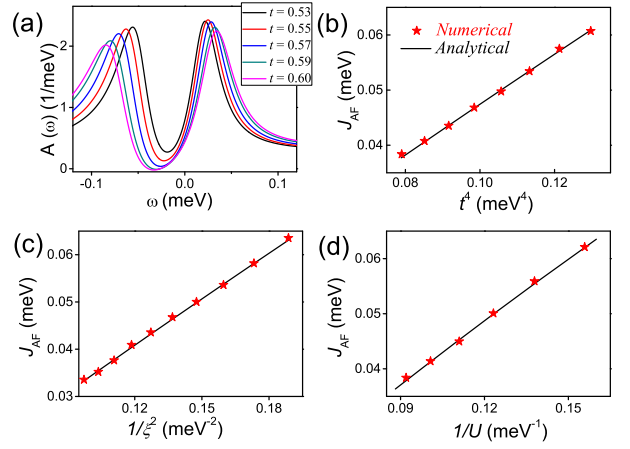}
\caption{(Color online)  The spectral function $A(\omega)=A_1 (\omega)=A_3(\omega)$ around $\omega=0$ is presented at $\varepsilon_2 = 2.0$ (in unit of meV) for different inter-dot coupling $t$ in (a). The dependence of the long-range antiferromagnetic exchange interaction $J_{\rm AF}$ on $t^4$, $1/\xi^2$ and $1/U$ are calculated in (b), (c) and (d), respectively. The other parameters are $U=1.2$  and $\xi=2.6$ in (b), $t=0.55$ and $U=1.2$ in (c), and $t=0.55$  and $\xi=2.6$ in (d).
The black lines and red scattered points  are calculated by analytic formula [\Eq{JAFA}] and numerical HEOM approach, respectively. The other parameters are the same as those in \Fig{fig2}.}
\label{fig4}
\end{figure}

In the limit of strong inter-dot couplings, e.g. $t=0.7$ meV, the overlapping of Kondo clouds of QD1 and QD3 becomes much stronger to induce a distinct peak at $\omega<0$ in $A_2(\omega)$, meanwhile the long-range $J_{\rm AF}$ has induced the phase transition from the degenerate KS state of individual QD to the LSS state (see \Fig{fig2}),  characterized by the splitting of the Kondo peaks of QD1 and QD3 in \Fig{fig3}(c). Since Kondo clouds are hard to be observed in experiments \cite{2009arx0911}, we suggest that they can be captured by their long-range overlapping or entanglement.

In the following, we will firstly derive an analytical expression of $J_{\rm AF}$ for isolated TQDs, and then prove that it is also valid in open-TQD system over a wide range of parameters.  We start from the Hamiltonian of \Eq{HS3} for isolated TQDs and constrain our derivation in the subspace with a total occupation number of $N_T\equiv N_1+N_2+N_3 =2$. The states of double occupation in QD2 (with zero occupation in QD1 and 3, i.e., the $|0,2,0\rangle$ states)
 are excluded because their energy is much higher than others. When $t=0$, $E_{|1,0,1\rangle}=2\epsilon_{1}, E_{|2,0,0\rangle}=E_{|0,0,2\rangle}=2\epsilon_{1}+U$ and  $E_{|1,1,0\rangle}=E_{|0,1,1\rangle}=2\epsilon_{1}+\xi$. Since only low-energy states are concerned, we substitute high-energy eigenvalues with their unperturbed ones and solve the secular equation to obtain the singlet and triplet states, where their splitting are defined as $J_{\rm AF}\equiv E_{T}-E_{S}$. After some algebra and assuming $t\ll U$, we get
\begin{align}
&E_{T}\approx2\epsilon_{1}+\frac{2t^{2}}{\xi};\\
&E_{S}\approx2\epsilon_{1}+\frac{2t^{2}}{\xi}-\frac{4t^{4}}{\xi^{2}U}.
\end{align}
Thus the $  J_{\rm AF}$ takes the form
\begin{equation}\label{JAFA}
  J_{\rm AF}\approx\frac{4t^{4}}{\xi^{2}U}.
\end{equation}
\Eq{JAFA} is similar to the antiferromagnetic exchange interaction in double QDs (DQDs) if one defines an effective next-neighbor inter-dot coupling between QD1 and 3 as $t'=t^2/\xi$ to rewrite \Eq{JAFA} as $ J_{\rm AF}\approx4t'^2/U$. It seems so, but a rigorous proof is required for open TQDs systems.  By investigating the splitting of Kondo peaks of QD1 or QD3, the features of $J_{\rm AF}$ can be studied, as that in DQDs systems \cite{2012prl266403}. The distance between the splitting peaks should equal to  $2J_{\rm AF}$.

The results of HEOM calculations on the formula of $J_{\rm AF}$ are summarized in \Fig{fig4}, where  a different $\varepsilon_2$ ( $\varepsilon_2=2.0$ meV) from that in \Fig{fig3} is chosen to verify the universality of our results.  The other parameters are the same as those in \Fig{fig2} unless specified otherwise. \Fig{fig4}(a) depicts $A(\omega)=A_1 (\omega)=A_3(\omega)$ around $\omega=0$ at different inter-dot coupling $t$ in the LSS phase at $t>0.5$ meV for $\varepsilon_2=2.0$ meV  (see \Fig{fig2}). As expected, the splitting of Kondo peak of QD1 or 3 is due to the antiferromagnetic exchange, thus $J_{\rm AF}$ can be well defined by the distance between two peaks. Then, the numerical $J_{\rm AF}$  as functions of $t^4$, $1/\xi^2$ and $1/U$ are respectively shown in \Fig{fig4} (b) to (d), together with the analytical results obtained from \Eq{JAFA} for comparison. In each figure, only one parameter varies and the others keep unchanged. One can see that almost all of the numerical data fall in the analytical lines in the range of parameters explored here. We thus conclude that  $J_{AF}\approx4t^{4}/\xi^{2}U$  is also valid in open-TQD systems over a wide range of parameters.

It is well acknowledged that the nearest-neighbor antiferromagnetic exchange can induce a  non-Fermi-liquid quantum critical point in DQDs, which is called `two-impurity problem'.  E. Sela and I. Affleck investigated the nonequilibrium transport through serial-coupled DQDs by bosonization and conformal field theory, and found a crossover from the critical point to the low energy Fermi liquid phase at finite temperature \cite{2009prl047201}.  For the DQDs, the crossover behavior of the quantum transition has been verified by the HEOM calculations on parallel-coupled DQDs \cite{2012prl266403}, as well as on serial-coupled ones (unpublished). For present TQDs study, we emphasize  some unique features shown in \Fig{fig4} as follows: ($i$)
The splitting peaks are not symmetric with respect to $\omega=0$ [see \Fig{fig4}(a)], which are different to the symmetric peaks in DQDs. The reason is that the electron-hole symmetry in QD1 and QD3 has been broken by the  overlapping of Kondo clouds [cf.~\Fig{fig3}]; ($ii$) The $t^{4}$-dependence shown in \Fig{fig4}(b) indicates that $J_{\rm AF}$ in TQDs is more sensitive to $t$, thus a larger $t$ is required to induce the phase transition from the KS to the LSS phase [cf.~\Fig{fig2}]; ($iii$) The $1/\xi^{2}$-dependence shown in \Fig{fig4}(c) suggests an easy way to manipulate $J_{\rm AF}$ in TQDs via gate control of the detuning energy. In this sense, the phase diagram shown in \Fig{fig2} is experimentally accessible.

\begin{figure}
\includegraphics[width=\columnwidth]{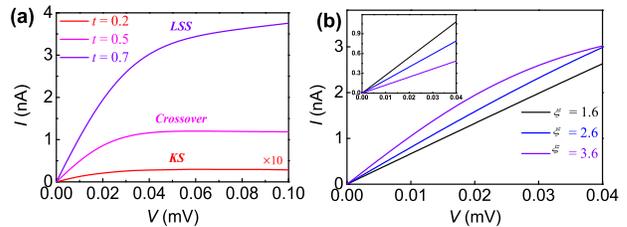}
\caption{(Color online) (a) The $I-V$ curves are shown at $\varepsilon_2 = 3.0$ (in unit of meV) for different phases:  KS ($t=0.2$), LSS ($t=0.7$) and the crossover region ($t=0.5$). (b) The $I-V$ curves are depicted for the TQDs system at various detuning energy $\xi$ in the LSS phase ($t=0.7$) at the temperature $K_{B}T = 0.03$ meV. Inset depicts the same curves  at  high temperature $K_BT = 0.3$.  The other parameters are the same as those in \Fig{fig2}.}
\label{fig5}
\end{figure}

In order to relate to experiments, we calculate the current-voltage ($I-V$) curves for $\varepsilon_2 = 3.0$ meV in different phases of KS ($t=0.2$ meV), LSS ($t=0.7$ meV) and the crossover region between them ($t=0.5$ meV). The results are shown in \Fig{fig5}(a), where the scale of the current of the KS phase is expanded by a factor of 10. A significant feature of \Fig{fig5} is the ``$J_{\rm AF}$-enhanced long-range transport'' in the LSS phase. At small $t$ (KS phase), although the Kondo peaks in QD1 and QD3 contribute a possible conductive channel, the high barrier of QD2 prevents the transfer of electrons from QD1 to QD3 [see \Fig{fig3}(a)]. As a consequence, the current remains at a small value in the entire bias range as shown in the figure. Increasing $t$ to $0.5$ meV drives the TQDs into the crossover region between KS and LSS. The overlapping of Kondo clouds begins to assist the tunneling of electrons from QD1 to QD3 through QD2. Accordingly, the current firstly increases near linearly with the bias voltage ($V<0.03$ mV) and gradually approaches a constant value of 1.2 nA ($V>0.04$ mV) in \Fig{fig5}(a).

The long-range transport is enhanced by $J_{\rm AF}$ in the LSS phase, manifesting it by enlarged current in \Fig{fig5}(a). The current at $t=0.7$ meV increases much faster than that at $t=0.5$ meV in the near-linear region at $V<0.03$ mV, with a slope of 100 nA/mV versus 40 nA/mV for the latter. Further increasing bias to $V>0.05$ mV will reveal more distinct differences between them. When it is saturated in the crossover region, the current in the LSS phase is not saturated and still increases in a near-linear way with another slope of 70 nA/mV. And the value of current in the LSS phase reaches 3.8 nA at $V=0.1$ mV, which is more than 100 times larger than that in the KS phase.

In order to highlight the effect of long-range entanglement of Kondo clouds  on the transport, we calculate the $I-V$ curves at various detuning energy $\xi$ at $t=0.7$ meV and summarize the results in \Fig{fig5}(b), where the insert  depicts the same curves  at  high temperature $K_BT = 0.3$ meV for a comparison. Ordinarily, the current would be decreased by higher tunneling barrier or increasing of  $\xi$, as shown in the insert.  However, an anomalous enhancement of current by higher barrier is clearly seen in \Fig{fig5}(b).
For example, the values of current at $V=0.02$ mV for $\xi=1.6$, $2.6$ and $3.6$ meV are respectively  $1.3$, $1.5$ and $2.0$ nA. The mechanism of the anomalous enhancement of current can be understood as follows:  As shown in \Fig{fig2}, with the increasing of $\xi$ at $t=0.7$ meV, the long-range correlation between QD1 and QD3 becomes stronger. At $\xi=3.6$ meV, the long-range overlapping or entanglement of the Kondo clouds is strong enough to form an extended conductive channel (see \Fig{fig3}). The electrons can thus transfer through QD1 to QD3. This kind of long-range transport with the aid of the overlapping of the Kondo clouds is obviously a many-body effect that is distinctly different from the low-order sequential tunneling and cotunneling. We comment that the anomalous enhancement of current  can be observed in routine experiments.

\section{Summary}
\label{Sum}

In summary, we have investigated the long-range interactions in triple quantum dots (TQDs) in the Kondo regime, by accurately solving the three-impurity Anderson  model with the hierarchical equation of motion (HEOM) formalism. For the occupation configuration of $(N_{1},N_{2},N_{3})=(1,0,1)$,  we demonstrate that there exists a long-range antiferromagnetic exchange interaction, $J_{\rm AF}$, which  can induce a continuous phase transition  from the separated Kondo singlet (KS) to the long-range spin singlet (LSS) state between edge dots. An expression of $J_{\rm AF}\approx4t^{4}/\xi^{2}U$ is analytically deduced  for isolated TQDs and numerically proved to be valid in open TQDs over a wide range of parameters. The long-range entanglement of Kondo clouds are quite robust at strong inter-dot coupling limit. Under equilibrium condition, it induces an unexpected peak in the spectral function of the middle dot whose singly occupied level keeps much higher than the Fermi level. Under nonequilibrium condition, higher inter-dot tunneling barrier induces an anomalous enhancement  of current. These novel features can be observed in routine experiments.

\acknowledgments

This work was supported by the NSF of China (No.\,11374363) and the Research Funds of Renmin University of China (Grant No. 11XNJ026). Computational resources have been provided by the Physical Laboratory of High Performance Computing at Renmin University of China. ZGZ is supported
by the Hundred Talents Program of CAS.


\end{document}